\documentclass[useAMS,usenatbib,a4paper]{mn2e}
\input{psfig}
\usepackage{amsfonts,amssymb,amsmath}
\usepackage{aas_macros}
\usepackage{times,varioref,color,multirow,textcomp}
\usepackage[
    dvips,
    a4paper=true,
    plainpages=true,
    pdfpagelabels,
    bookmarks=true,
    bookmarksopen=false,
    bookmarksopenlevel=2
    bookmarksnumbered=true,
    bookmarkstype=toc,
    colorlinks=true,
    citecolor=red,
    linkcolor=blue,
    menucolor=green,
    urlcolor=magenta,
]{hyperref}
  
\def \Mpch{h^{-1}{\rm Mpc}}
\def \Msun{{\rm M}_\odot}

\begin{document}


\title[Galaxy clustering in WMAP cosmologies]
      {Exploring Galaxy Formation Models and Cosmologies with Galaxy Clustering}
\author[Kang {\it et al.}]
       {X. Kang$^{1}$\thanks{E-mail:kangxi@pmo.ac.cn}, M. Li$^{1}$, W.P. Lin$^{2}$, P.J. Elahi$^{2}$\\
        $^1$The Purple Mountain Observatory, 2 West Beijing Road, Nanjing 210008, China\\
        $^2$Key Laboratory for Research in Galaxies and Cosmology, Shanghai Astronomy Observatory, Nandan Road 80, Shanghai 200030, China}


\date{}

\pagerange{\pageref{firstpage}--\pageref{lastpage}}
\pubyear{2011}

\maketitle

\label{firstpage}


\begin{abstract}
Using N-body simulations and galaxy formation models, we study the galaxy stellar mass correlation and the two-point auto-correlation. The simulations are run with cosmological parameters from the WMAP first, third and seven year results, which mainly differ in the perturbation amplitude of $\sigma_{8}$. The stellar mass of galaxies are determined using either a semi-analytical galaxy formation model or a simple empirical abundance matching method. Compared to the SDSS DR7 data at $z=0$ and the DEEP2 results at $z=1$, we find that the predicted galaxy clusterings from the semi-analytical model are higher than the data at small scales, regardless of the adopted cosmology. Conversely, the abundance matching method predicts good agreement with the data at both $z=0$ and $z=1$ for high $\sigma_8$ cosmologies (WMAP1 \& WMAP7), but the predictions from a low $\sigma_8$ cosmology (WMAP3) are significantly lower than the data at $z=0$. We find that the excess clustering at small-scales in the semi-analytical model mainly arises from satellites in massive haloes, indicating that either the star formation is too efficient in low-mass haloes or tidal stripping is too inefficient at high redshift. Our results show that  galaxy clustering is strongly affected by the models for galaxy formation, thus can be used to constrain the baryonic physics. The weak dependence of galaxy clustering on cosmological parameters makes it difficult to constrain the WMAP1 and WMAP7 cosmologies.

\end{abstract}




\section{Introduction}
\label{sec:intro}
The Cold Dark Matter (CDM) paradigm successfully describes the formation of structure in the Universe. In this paradigm, galaxies form in the potential wells of dark matter haloes via gas cooling and subsequent star formation \citep{white1978}. Observations, such as satellite kinematics \citep{conroy2007,more2009}, galaxy-galaxy lensing (e.g., \citealp{mandelbaum2006}) and galaxy clustering (e.g., \citealp{zehavi2005}), have shown that the properties of galaxies, such as their stellar mass, colour and morphology, are closely related to the inferred mass of their host haloes. Thus any successful model for galaxy formation must reproduce these observations.

\par 
Here we focus on the observed clustering of galaxies and explore the effectiveness of this quantity in constraining galaxy formation models and cosmological parameters as this clustering is very sensitive to the host halo mass in which galaxies live. The two-point correlation function (2PCF) of galaxies has been accurately measured by large surveys, such as the Two Deep Field Galaxy Redshift Survey \citep{colless2001}, the Sloan Digital Sky Survey \citep{york2000} at $z=0$, and the DEEP2 Galaxy Redshift Survey at $z=1$ \citep{davis2003}. The 2PCF appears to be a power-law over a wide range of scales, and it depends on redshift, luminosity, colour and morphology of galaxies \citep{norberg2002,coil2004,zehavi2002,zehavi2005,li2006}. Recently, \cite{li2009} extended the traditional 2PCF by measuring the stellar mass correlation function from the SDSS DR7 data, the so-called stellar mass correlation function (SMCF). This quantity provides additional constraint on galaxy formation models, as it depends on the relative mass of galaxies at given scales.

\par
One tool often used to study galaxy clustering is Semi-Analytical Models (SAMs) of galaxy formation \citep{white1991}. Early studies found that these models predicted clustering which marginally agreed with the data (e.g., \citealp{kauffmann1999}). Recent deep surveys, which have included more faint galaxies, have shown that currently SAMs predict too high a clustering amplitude at small scales \citep{weinmann2006,li2007}. The excess clustering in these studies was primarily due to the over-abundance of faint galaxies in the models. The recent model of \cite{guo2011} removed this over-abundance and was able to reproduce the local Stellar Mass Function (SMF) down to very low mass end. However, this model still over-predicts the small-scale clustering, albeit to a smaller degree. These results indicate that it is not solely abundance of galaxies but their spatial locations which lead to a higher clustering at small scales. \cite{guo2011} suggested that perhaps a low $\sigma_{8}$ universe might remove this discrepancy. 

\par
Another way to model galaxy clustering is the Abundance Matching Method (AMM, e.g. \citealp{vale2004}). This method assumes that there is a monotonic relation between a galaxy's stellar mass and its host halo's mass (or progenitor host halo's mass at the time of its accretion for a subhalo containing a satellite galaxy), and the stellar mass can be obtained by matching the halo (or subhalo) abundance  to the observed SMFs. The biggest advantage of this approach is that the observed SMF is perfectly reproduced. It was found that this simple approach can well reproduce the properties of galaxy clustering seen in the SDSS at $z=0$ (e.g., \citealp{moster2010}). Extensions of this simple method by including more complicated dependence on galaxy type, redshift, host halo mass can be found elsewhere \citep{wang2006,behroozi2010,neistein2011}. 

\par
In this paper, we use the SAM of \cite{kang2008} (hereafter K08, model I) and the AMM (model II) to produce galaxy catalogues in three cosmologies based on the Wilkinson Microwave Anisotropy Probe (WMAP) first year, third year and seventh year results (\citealp{spergel2003}: WMAP1; \citealp{spergel2007}: WMAP3; \citealp{komatsu2011}: WMAP7). The model of \cite{kang2008} slightly over-predicts the abundance of low-mass galaxies, and the parameters were tuned to fit the SMFs of \cite{cole2001} and \cite{bell2003}. Here we modify it slightly by extending the gas cooling time in low-mass haloes to better match the low-mass end of the local SMF of \cite{li2009}. By constructing a mock galaxy catalogue, we are able to determine what galaxies contribute to the small-scale clustering. We also investigate whether the current favored cosmological parameters from the WMAP results, especially $\sigma_{8}$, can be better constrained.

\par 
The paper is organized as following: In Section~\ref{sec:method}, we outline how we construct the galaxy catalogue using merger trees from N-body simulations and the two methods discussed previously in the text, SAMs and the AMM. In Section~\ref{sec:results}, we present the predictions for the stellar mass clustering, 2PCF and their dependence on mass and colour. We focus on the differences between the two models and investigate the origin for the discrepancy between SAM and the data observed in previous studies. Finally, we conclude with a summary and discussion of our results in Section~\ref{sec:cons}. 

\section{Building the Mock Galaxy Catalogues}
\label{sec:method}
To predict the 2PCF and the stellar mass correlation, we need to produce a large sample of galaxies in a large cosmological volume. To achieve this, we build galaxy catalogues using N-body simulations of three different cosmologies. For each model galaxy, its stellar mass is determined in two ways. Model I is the slightly modified semi-analytical model of K08, which is based on \cite{kang2005,kang2006}. This model self-consistently models the physical processes governing stellar mass evolution, such as gas cooling, star formation, supernova and AGN feedback. Model II is the abundance matching method, which determines the stellar mass of each galaxy by using the observed SMF. These two methods are discussed in detail below.

\subsection{Simulations \& halo merger trees}
For both models we must first extract haloes from N-body simulations and determine their accretion history using a merger tree in order to ``paint'' luminous matter on to them. The N-body simulations were performed using the Gadget-2 code \citep{springel2005}. The three cosmologies are based on the WMAP1, WMAP3, and WMAP7 results, which mainly differing in the amplitude of power spectrum with $\sigma_{8}$. The amplitude of $\sigma_{8}$ for the WMAP1, WMAP3, and WMAP7 are $0.9$,$0.73$ and $0.8$ respectively. From here on, we will refer WMAP1 and WMAP7 as high$-\sigma_8$ cosmologies and WMAP3 as a low$-\sigma_8$ cosmology. All simulations are run with $1024^{3}$ dark matter particles in a cube of 200~$\Mpch$ on each side.

\par
We briefly discuss the construction of halo merger trees here, for a more detailed discussion see \cite{kang2005}. At each snapshot, dark matter haloes are identified using the Friends-of-Friends (FOF) algorithm. For each FOF halo we determine its virial radius, $r_{\rm vir}$, defined as the radius centered on the most bound particle inside of which the average density is $\Delta_{c}(z)$ times the average density of the universe \citep{bryan1998}. The mass inside $r_{\rm vir}$ is called as the virial mass $m_{\rm vir}$. Inside each FOF halo, subhaloes are identified using the SUBFIND \citep{springel2001}. Subhaloes are relics of haloes that have been accreted by a larger halo. Using the FOF and subhalo catalogue, we can construct the (sub)halo merger trees and produce model galaxies inside each (sub)halo.

\par
The galaxies produced in this merger tree are called central galaxies if the galaxy is the largest galaxy at the center of each FOF halo. Any other galaxies associated with subhaloes are called satellites. For each satellite galaxy, we trace back to the time when it was a central galaxy of a FOF halo, and label the virial mass of the FOF halo as $M_{acc}$. For central galaxy, the mass $M_{acc}$ is the current virial mass of its FOF halo. This mass, $M_{acc}$, will later be used to determine the stellar mass in model II.

\subsection{Modeling galaxy stellar masses}
\label{sec:method1}
\begin{figure}
 \centerline{\psfig{figure=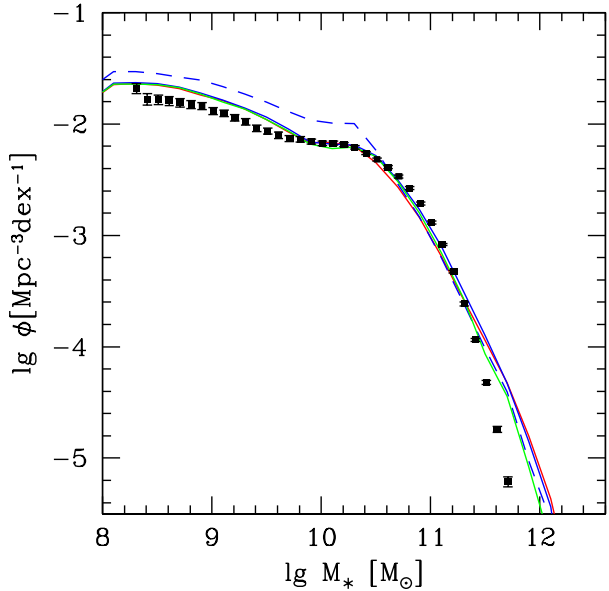,width=0.45\textwidth}}
 \caption{The local stellar mass functions. The data points shows the \protect\cite{li2009} results measured from the SDSS DR7. The red, blue and green solid lines are the predictions from model I for the WMAP1, WMAP7 and WMAP3 cosmologies with $f_c=0.15$. The dashed line is the result of K08 ($f_c=1.0$) for the WMAP7 cosmology. Note that the turn off at lower mass ($< \sim 10^{8}M_{\odot}$) is due to the resolution of our simulations.}
 \label{fig:SMF}
\end{figure}

The main ingredients of the SAM used in this study are described in detail in \cite{kang2005}. These models allow one to easily investigate how galaxy properties vary as the underlying assumptions regarding the baryonic physics are changed (e.g.,~\citealp{bower2006,delucia2007,somerville2008,guo2011}). Although SAMs successfully reproduce a wide range of observations, they have difficulty reproducing the shallow slope of SMF at the low-mass end. Recently \cite{guo2011} found that an enhanced supernova feedback and longer gas reincorporation time could reproduce the shallow slope measured from the SDSS DR7 by \cite{li2009} down to a stellar mass of $10^{8}M_{\odot}$. 

\par
The model of \cite{guo2011} has the effect of decreasing the gas cooling rate, especially in low-mass haloes. Here we introduce a minor modification to the model of K08, to obtain a better match to the SMF. We parameterize the gas cooling in low mass haloes as,
\begin{equation}
\dot{M}_{cool} = f_{c}*m_{hot}/t_{dyn}
\end{equation}
where $\dot{M}_{cool}$ is the gas cooling rate onto the central galaxy in a halo, and $m_{hot}$ is the total gas content in that halo, and $t_{dyn}$ is the dynamical time of the halo. In the K08 model, $f_{c}$ is effectively set to $1$ which resulted in a steep slope at low-mass end. The parameters in K08 were normalized using the SMFs of \cite{cole2001} and \cite{bell2003}. In this paper, we tune our model parameters to best match the SMF of \cite{li2009}, which change the parameters only slightly. 

\par
In Fig.~\ref{fig:SMF}, we show the SMFs from our models and \cite{li2009}. Overall the observed SMF is well reproduced, though our models have an over abundance of massive galaxies ($M_{\ast} > 3\times 10^{11}M_{\odot}$). We will later see that this excess of massive galaxies is not the reason for the over-prediction of the SMCF. The solid lines are  results with $f_c=0.15$ for the three cosmologies used in the paper, and the dashed line is the K08 result with $f_c=1.0$ for the WMAP7 cosmology.  Compared to the K08 result, it is found that a lower $f_c$ can reduce the slope at small scales. Lowering $f_{c}$ further will decrease the faint-end slope, but it will also under-predict the abundance of galaxies around the characteristic stellar mass, $M_{\ast}$. Note that in our model gas which is heated by supernova feedback is ejected from the central galaxy, but remains in the halo. In the model of \cite{guo2011}, the fraction of gas ejected depends on the halo's potential, and this ejected gas is re-incorporated over longer time scales. This process is effectively included here by using $f_c <1$. Note that at masses lower than $\sim 10^{8}\Msun$, our results are affected by the mass resolution of our simulations, which artificially reduces the number of low mass galaxies. 

\par
The second method we use to determine stellar mass is the AMM, originally proposed by \cite{vale2004}. Observations have shown that there is a tight scaling relation between galaxy properties and the host halo mass (e.g., \citealp{mandelbaum2006}). By assuming that there is a monotonic relation between galaxy stellar mass and its host halo mass at accretion ($M_{acc}$), the stellar mass $M_*$ can be determined by matching the mass function of $M_{acc}$ of the model galaxy to the observed SMF, i.e.~$n(>M_{acc})=n(>M_*)$. This match leaves a relation between the stellar mass and halo mass, called as $M_{\ast}-M_{h,acc}$ relation. Note that in this paper, we do not include any scatter in the $M_{\ast}-M_{h,acc}$ relation and also neglect any possible evolution with redshift. A full accounting of these effects is beyond the scope of this paper. We refer interested readers to \cite{moster2010} and \cite{behroozi2010} for discussions of these uncertainties. 

\par
\begin{figure}
 \centerline{\psfig{figure=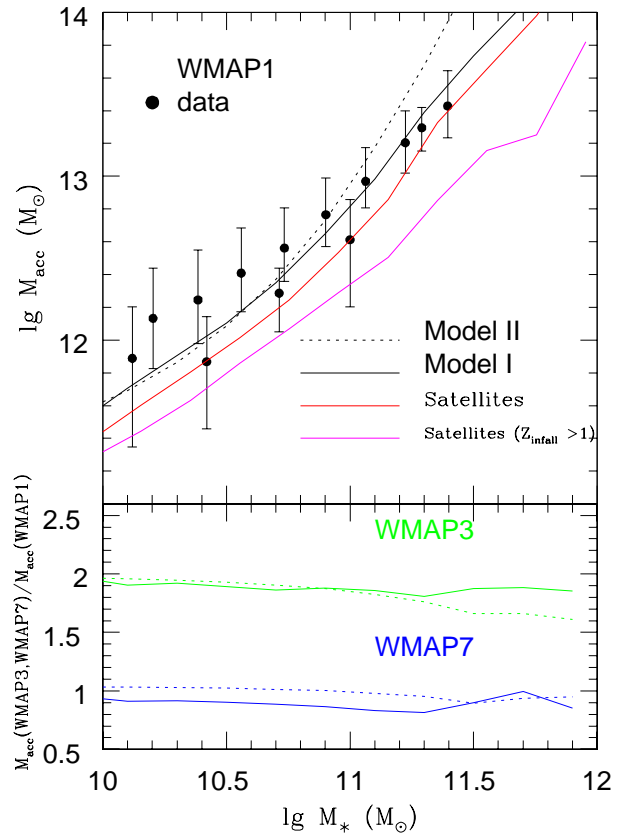,width=0.45\textwidth}}
 \caption{The stellar mass and halo mass ($M_{\ast}-M_{h,acc}$) relation. Upper panel: the data points are from observation of \protect\cite{mandelbaum2006} and \protect\cite{more2009}, and the lines are predictions for the WMAP1 cosmology, with dotted lines for model II and others for model I. Lower panel: the predicted relation for other cosmologies, as shown by their ratio to the WMAP1 results. Here we have shifted the WMAP3 results by 1 for clarity.}
 \label{fig:ms-macc}
\end{figure}

In Fig.~\ref{fig:ms-macc}, we show the matched $M_{\ast}-M_{h,acc}$ relation. The upper panel shows the relations from the WMAP1 cosmology, and the lower panel are for the WMAP3 and WMAP7 results, shown by their ratio to the WMAP1 one. Note that the WMAP3 results are shifted by 1 for clarity. The solid lines are predictions from model I and dotted lines are the matched one from model II. In the upper panel, the observational results are from \cite{mandelbaum2006} and \cite{more2009}. Also shown are the relations for all satellite galaxies (red line), and satellites accreted at redshift $z_{acc} >1$ (magenta line), which we will now on refer to as early accreted satellites. 

\par
Clearly, the two model predictions are very similar to the observations except that model II over-predicts the halo mass for massive galaxies. In model I, such a $M_{\ast}-M_{h,acc}$ relation depends on whether a galaxy is a central galaxy or a satellite. For satellite galaxies, especially early accreted satellites, their host halo's mass at the time of accretion is lower relative to central galaxies. This is because the gas cooling and star formation efficiencies are  higher at high redshift in the model. We will later see that these early accreted satellites now live in massive haloes, and it leads to a higher clustering at small scales. The lower panel shows that the predicted $M_{\ast}-M_{h,acc}$ relation has little dependence on cosmology as we have set the model parameters to best fit the observed SMF. The resulting  star formation efficiencies in given halo mass do not vary drastically between the three cosmologies analyzed. 

\section{Galaxy clustering}
\label{sec:results}
\begin{figure*}
 \centerline{\psfig{figure=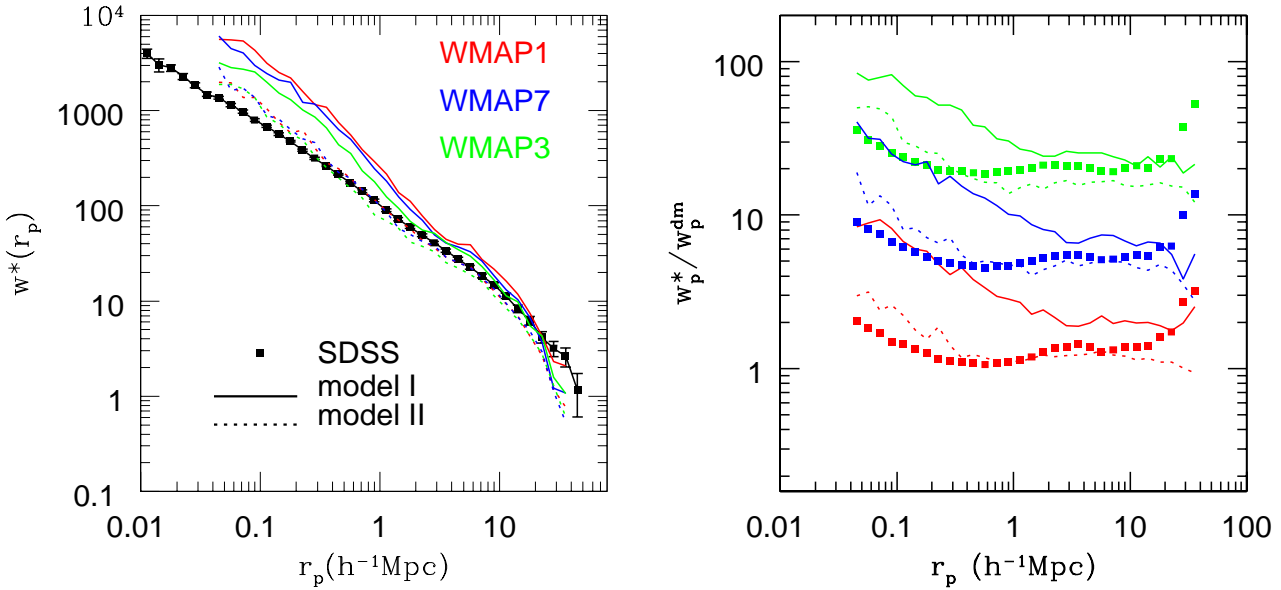,width=0.95\textwidth}}
 \caption{The projected stellar mass correlations. Left panel: data points are from \protect\cite{li2009}. The colour lines are  predictions from Model I (solid lines) and Model II (dotted lines), with red, blue and green colours referring to the WMAP1, WMAP7 and WMAP3 cosmologies. Right panel: the galaxy bias relative to the dark matter particles. Note that the results for the WMAP3 and WMAP7 cosmologies are shifted by  factors of 10 and 3 for clarity.}
 \label{fig:SMC}
\end{figure*}

\begin{figure*}
 \centerline{\psfig{figure=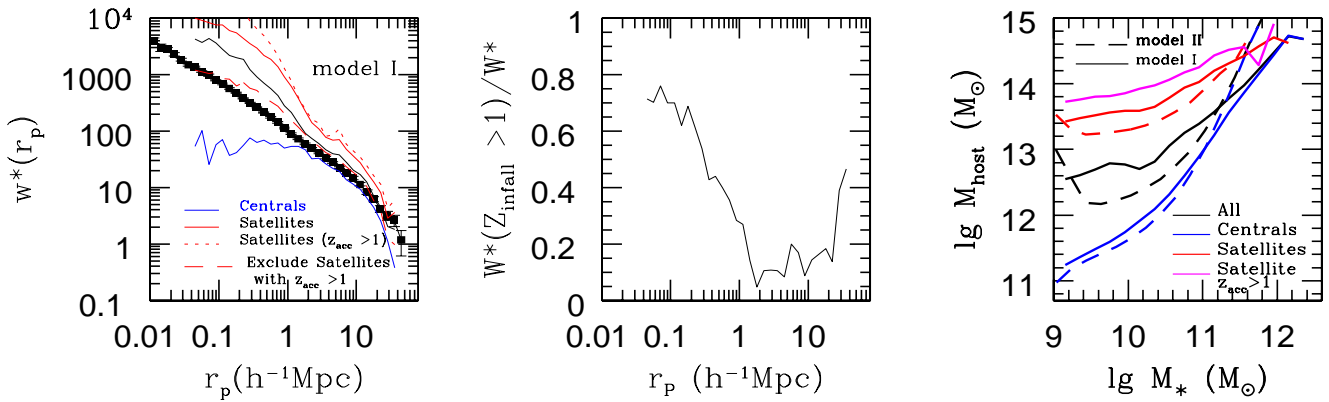,width=0.99\textwidth}}
 \caption{The projected stellar mass correlations with dependence on galaxy type. Here only results from the WMAP1 cosmology are shown. Left panel: predictions from model I for central and satellite galaxies. Middle panel: the contribution of early accreted satellites ($z_{acc}>1$) to the total clustering. Right panel: the host halo mass of galaxies at $z=0$ in the two models. }
 \label{fig:SMC-type}
\end{figure*}

The traditional 2PCF only uses the position of each galaxy, and is thus incapable of constraining the properties of galaxy pairs at given distance. \cite{li2009} extended it by weighting each galaxy with its stellar mass, namely the stellar mass correlation function (SMCF). The projected SMCF is written as,
\begin{equation}
 w_{p}^{*}(r_{p}) = \int_{-\pi_{max}}^{+\pi_{max}} \xi^{*}(r_{p}, \pi) d\pi,
\end{equation}
where $\xi^{*}(r_{p},\pi)$ is the projected redshift-space correlation weighted by the product of the stellar masses in each pair, with distance of $r_{p}$ and $\pi$. We compute the predicted SMCFs from the models in the same way as \cite{li2009}.

\par
In the left panel of Fig.~\ref{fig:SMC} we show the projected stellar mass correlation. The data points show the measurements of \cite{li2009} from the SDSS DR7. The solid, dotted lines show the predictions from model I and model II, with red, blue and green colour referring to the WMAP1, WMAP7, and WMAP3 cosmologies respectively. It shows that model I over-predicts the SMCFs in all three cosmologies studied. The size of the discrepancy depends on the mass scale. At small masses the prediction is higher by a factor of $2-5$ but at large masses the predicted SMCF is only $30\%$ higher. The coloured lines show that lower $\sigma_{8}$ only slightly decreases the clustering amplitudes. The predictions from model II agree better with the data, from large scales to $0.3~\Mpch$. At small scales, the predictions are only higher than the data by $\sim20\%$.

\par
In the right panel of Fig.~\ref{fig:SMC}, we show the galaxy bias, defined as the ratio of galaxy clustering relative to the clusering of underlying dark matter distribution, $w^{\ast}_{P}/w^{dm}_{P}$, where $w^{dm}_{P}$ is measured directly from our simulations.  Note that the results from the WMAP7 and WMAP3 cosmologies are shifted for clarity. On large scales, results from model II in the high $\sigma_8$ cosmologies fit the data equally well, but the WMAP3 cosmology is systematically lower. \cite{li2009} use the shape of galaxy bias to constrain cosmology with the assumption that the galaxy bias is flat on large scales, and rising at small scales based on the model of \cite{delucia2007}. Indeed, we find that both models predict similar shapes, indicating that the shape of bias is a generic prediction of galaxy formation models.
\begin{figure*}
 \centerline{\psfig{figure=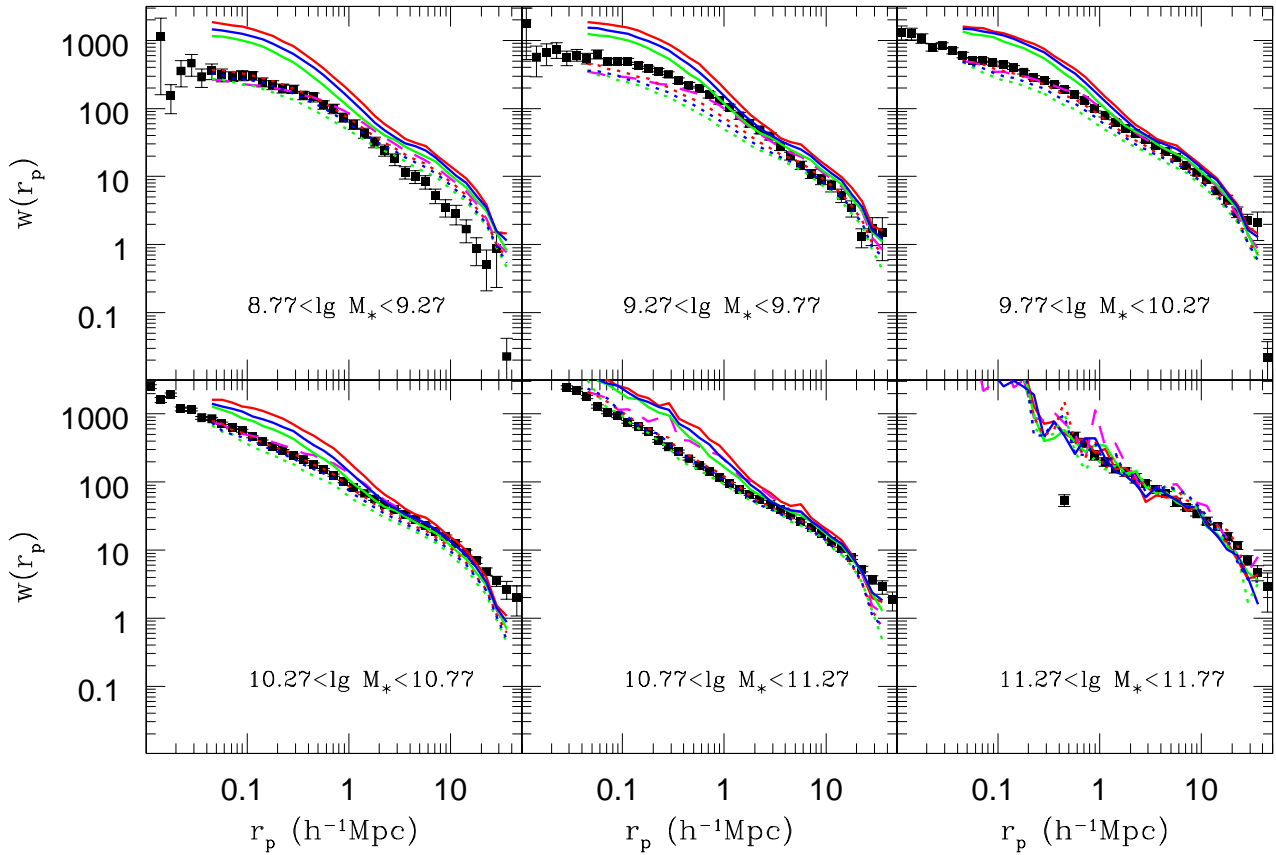,width=0.95\textwidth}}
 \caption{The projected 2PCF of galaxies in different stellar mass bins. Data points are from Li \& White (2009). Solid lines are for model I and dotted lines for model II. The dashed magenta lines show results from model I with the exclusion of early accreted satellites ($z_{acc}>1$).}
 \label{fig:2pcf-ms}
\end{figure*}

\par
To further explore the origin of the discrepancy of model I with the data, we show in Fig.~\ref{fig:SMC-type} the clustering of galaxies of different types. The left panel shows the SMCFs of centrals, satellites and early accreted satellites, respectively. The middle panel gives the contribution of early accreted satellites to the total SMCF. In the right panel we show the host halo mass at $z=0$ for galaxies with different types, also shown are the results from model II (dashed lines).

\par
The left and middle panels show that satellites, primarily early accreted satellites, are the main contribution to the clustering amplitude at small scales. By excluding these early accreted satellites, the clustering amplitude can be significantly suppressed at small scales (dashed line). The right panel shows that  satellites primarily live in big haloes ($\gtrsim10^{13}~\Msun$) and those early accreted ones ($z_{acc}>1$) reside in even more massive haloes $(M_{\rm host}\gtrsim10^{14}~\Msun)$. 

\par
The results in the above figure are easy to understand. In the CDM universe, massive haloes grow by merger of small haloes at early times, and N-body simulations have shown (e.g., \citealp{gao2004}) that  massive haloes accrete more low-mass haloes. The magenta line in Fig.~\ref{fig:ms-macc} implies that star formation efficiency in low-mass haloes from model I is higher than that from model II.  Consequently, for given stellar mass, the galaxies in model I live in more massive haloes.  From the halo models (e.g., \citealp{mo1996}; \citealp{cooray2002}), it is known that massive haloes are strongly clustered at all scales. The higher SMCFs from model I imply that the stellar mass of satellites are over-estimated, indicating that in the SAM either the star formation efficiency in low-mass haloes was too high at high redshift, or the tidal stripping is too inefficient for satellites.

\par
In Fig.~\ref{fig:2pcf-ms}, we show the projected 2PCFs of galaxies in different stellar mass bins for the two models. As in previous plots, the solid lines are for model I and dotted lines for model II. It is found that model I agrees with the data on all scales for very massive galaxies ($> 10^{11.27}~\Msun$), but for less massive galaxies, the predictions agree with the data only on large scales. For the lowest mass bin ($\lg M_{\ast} < 9.27$) the predictions are higher on all scales. As noted by \cite{guo2011}, the low-mass galaxies in the SDSS DR7 are severely affected by a few structures and distorted by peculiar velocities. Accounting for these effects is beyond the scope of our paper, so we limit our discussion on galaxies with mass higher than $10^{9.77}~\Msun$.

\begin{figure*}
 \centerline{\psfig{figure=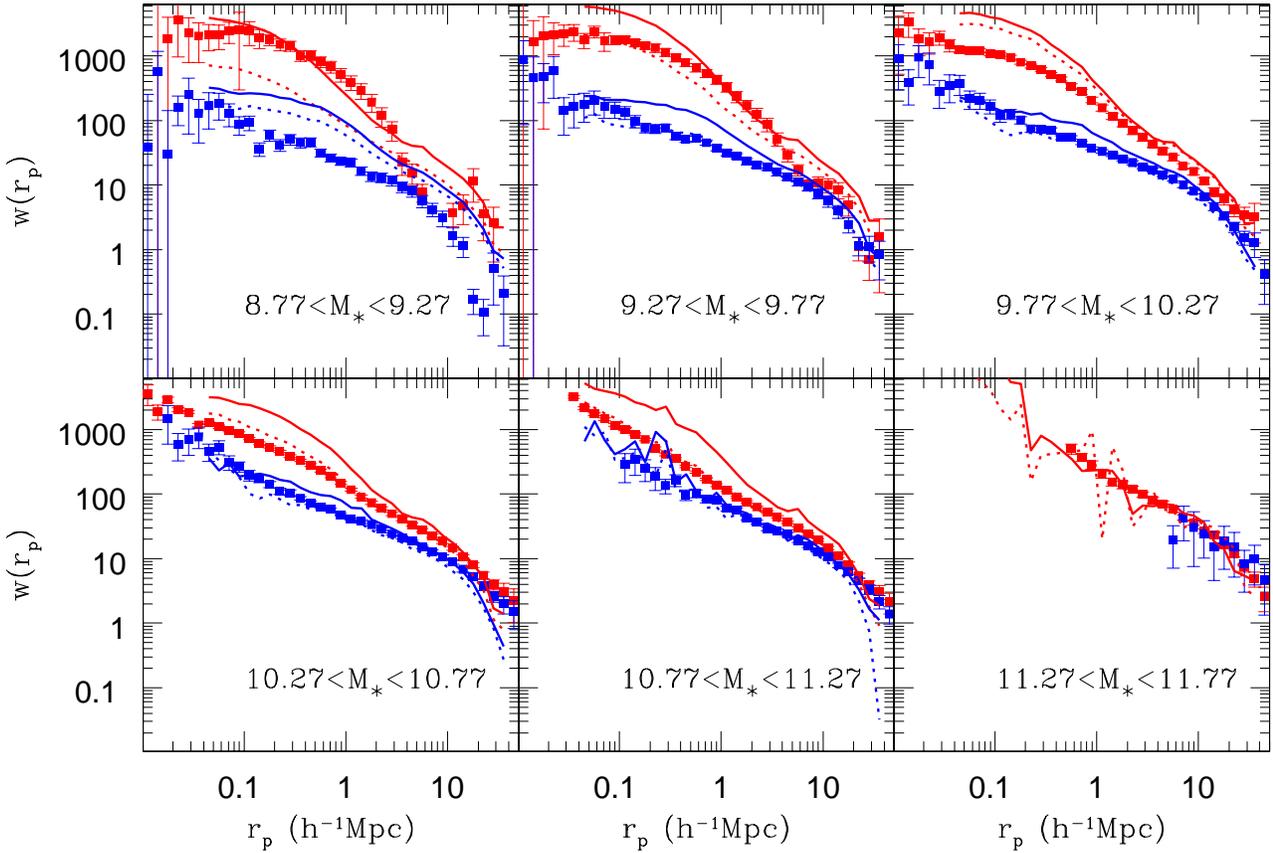,width=0.95\textwidth}}
 \caption{The projected 2PCFs of red and blue galaxies in different stellar mass bins. The solid lines are for model I and dotted ones for model II, with red and blue lines for red and blue galaxies, respectively. For clarity here only predictions from the WMAP1 cosmology are shown, as the other two cosmologies give similar results.}
 \label{fig:2pcf-colour}
\end{figure*}

\begin{figure*}
 \centerline{\psfig{figure=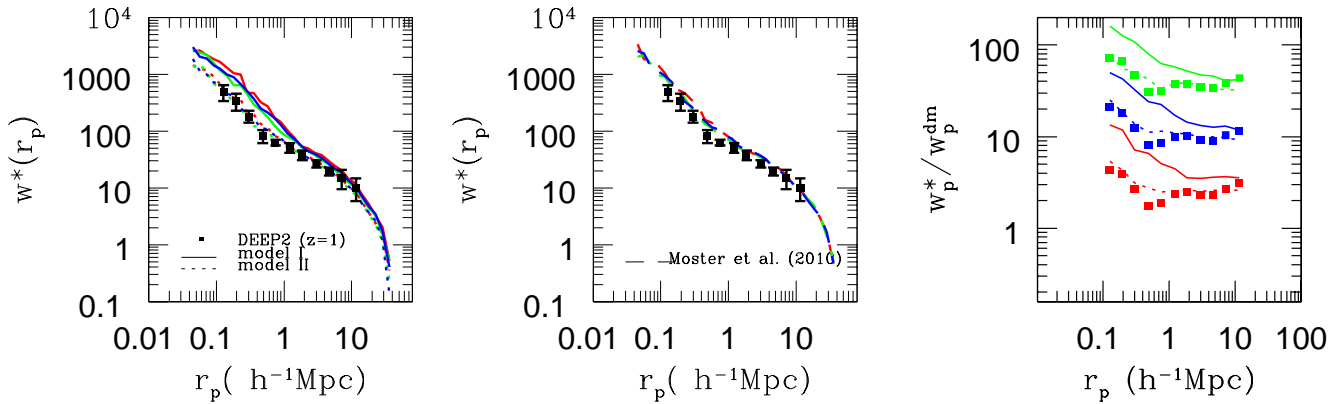,width=0.99\textwidth}}
 \caption{The stellar mass correlation function at $z=1$. The right panel is for the bias. The predictions from model II have used the local $M_{\ast}-M_{h,acc}$ relation (dotted line in the left panel) and the evolved one from \protect\cite{moster2010} (dashed lines in the middle panel). }
 \label{fig:SMC-z1}
\end{figure*}

\par
Remarkably, the predictions from model II are lower than that from model I and agree well with the data on all scales for galaxies with mass larger than $10^{9.77}~\Msun$. Such an agreement was also recently shown by other studies (e.g., \citealp{moster2010}). The most distinct improvement from model II  is the suppression of clusterings on small scales. We have shown that this is because galaxies are on average living in low-mass haloes in model II. As a test, we show predictions  with the exclusion of early accreted satellites ($z_{acc}>1$) in model I using the dashed magenta lines. As seen from Fig.~\ref{fig:SMC-type} the early accreted satellites dominate the power on small scales, neglecting them significantly improves the agreement between model I and the observations.

\par
The excess of clustering on small scales from the SAMs has already been well known. It was recently suggested that a low $\sigma_{8}$ universe may provide better fit to the clustering on small scales, as such a cosmology contains fewer massive haloes (e.g., \citealp{guo2011}). Here we find that predicted galaxy clustering from the SAM is only slightly suppressed in the low$-\sigma_{8}$ (WMAP3) universe and still lies above the data. Actually, we find that using model II with a high $\sigma_8 \sim 0.9$ (WMAP1) cosmology provides slightly better fit to the data than the other two cosmologies. Our results indicate that the model of determing  galaxy mass plays the dominate role in the predicted clusterings. Consequently, due to our poor understanding of galaxy formation physics, we must rely on other observations, such as cosmic microwave background or weak lensing, which have little dependence on small-scale baryonic processes, to provide tight constraints on cosmological parameters.

\par
Now we consider the dependence of galaxy clustering on colour. The unobscured galaxy color is calculated using the luminosity in multiple wavebands. These luminosities are calculated by combining the star formation history with the stellar population synthesis models, and both are self-consistently modeled in our model I. We classify a galaxy as red or blue using the color-magnitude relation from \cite{li2006}. As model II does not predict galaxy colour and to study the colour dependence  in this model, we make a naive assumption that the galaxy has the same color as its counterpart in model I. This assumption is probably true as we know that a galaxy's colour is mainly affected by the merger history of its host halo. For example, most satellites in model I are red because they were accreted at early times and their current star formation rates are very low. These objects are also expected to be red in model II as their stellar mass is set by the host halo mass at accretion, which implies that satellites experience no star formation once they are accreted. Most centrals are blue in model I as they undergo continuous gas cooling and star formation. They are also expected to be blue in model II as their stellar mass are determined by the current halo mass, again imply continuous star formation (see the good agreement of $M_{s}-M_{h,acc}$ relation in Fig~\ref{fig:ms-macc} for centrals in both models). 

\par
In Fig.~\ref{fig:2pcf-colour} we show the predicted 2PCFs from model I (solid lines) and model II (dotted lines). Here only results from the WMAP1 cosmology are shown as the other cosmologies produce similar results. The red and blue lines are for red and blue galaxies, respectively. Quantitively, model I reproduces the colour dependence seen in the data, that is red galaxies in each mass bin are more clustered than blue ones. The predicted clustering of blue galaxies is in better agreement with the data than that of red galaxies. Model II provides better match to the data  due to the suppression of clustering of red galaxies. The fact that we can reproduce the colour dependence indicates that the SAMs have correctly modeled the main physics governing galaxy colour, such as AGN feedback in massive centrals and tidal strangulation of hot halo gas from satellites.  

\subsection{Stellar mass clustering at $z=1$}
From the above, we have seen that model II provides better match to the data than model I at $z=0$. Here we investigate how the predictions change with redshift, and in particular, we check if the clustering properties at $z=1$ can place constraints on the used cosmologies. To obtain the stellar mass of model galaxy in model II at $z=1$, we use both the local $M_{\ast}-M_{h,acc}$ relation in Fig.~\ref{fig:ms-macc}, and the one given by \cite{moster2010} who obtained it by fitting the SMFs at $z=1$. The results are shown in Fig.~\ref{fig:SMC-z1}, where the data points of the DEEP2 are from \cite{li2011}. 

\par
Seen from the solid lines in the left panel that the SMCF from model I is still excessively clustered on small scales, and a low $\sigma_{8}$ universe still can not resolve this discrepancy. The dotted lines in the left panel shows that model II with the local $M_{s}-M_{h,acc}$ relation can well reproduce the clustering at all scales, and the predictions are identical for all the three cosmological models. Using the $M_{s}-M_{h,acc}$ relation of \cite{moster2010} increases the clustering slightly but the results are still acceptable. This suggests that the stellar mass to halo mass relation is almost in place at $z=1$, indicating that the tidal stripping effect for satellites is not significant at $z<1$. The right panel shows the stellar mass bias, which again has a smooth transition that it is flat on large scales and growing on small scales, similar to the one in Fig.~\ref{fig:SMC}. This behaviour is independent of redshift, further supporting the conclusion of \cite{li2009} that this form of galaxy bias is a generic prediction of galaxy formation models.

\section{Conclusions and Discussions}
\label{sec:cons}

In this paper, we have studied the stellar mass clustering and the projected two-point correlation functions of galaxies, and compared the model predictions to data from the SDSS $(z=0)$ and the DEEP2 $(z=1)$ \citep{li2009,li2011}. In order to explore the feasibility of constraining cosmological parameters using these quantities, we ran N-body simulations with three different cosmologies based on the WMAP1, WMAP3, WMAP7 results, which mainly differ in $\sigma_{8}$ values ($0.9, 0.73, 0.81$ respectively). We then populated the simulations with model galaxies, and determine their stellar mass using two different approaches: model I is a the Semi-Analytical Model, which self-consistently models the physics of star formation, model II is an empirical model where stellar masses are obtained by matching the observed stellar mass function. We have obtained a few interesting results listed below. 

\begin{itemize}
 \item The stellar mass clustering function predicted by the semi-analytical model is excessively clustered on small scales, and is still about $30\%$ higher even at larger scales. The projected two-point correlation agrees with the data on large scales but is still too high on small scales. These results imply that the excess clustering is from an over-prediction of the stellar mass of satellites in massive haloes. We further found that this excess is mainly from  satellites accreted at early times. These galaxies at $z=0$ live in massive haloes larger than $10^{14}M_{\odot}$ and as a result are strongly clustered. We also found that a low $\sigma_{8}$ universe ($\sigma_{8}=0.8, 0.73$) will only slightly decrease the small-scale clustering, but the discrepancies with observational data remain. 

 \item The abundance matching model provides a much better fit to the data at both $z=0$ and $z=1$ than the semi-analytical model. This improvement is primarily due to the suppression of the number of low mass satellite galaxies residing in massive haloes. We found that the WMAP1 and WMAP7 cosmologies ($\sigma_{8}=0.9, 0.8$) both provide acceptable fit to the data, but the WMAP3 cosmology ($\sigma_{8}=0.73$) predicts a galaxy clustering on large scales at $z=0$ that is lower than the observed one.

 \item Qualitatively, the colour dependence of clustering is reproduced from model I, with the over-prediction of clustering of red galaxies on small scales. The predicted clusterings of blue galaxies agree better with the data. By suppressing the stellar mass of satellites in model II, the colour dependence of clustering is well reproduced in this model. These results indicate that the SAMs can marginally capture the main physics governing galaxy colour. Here we note that the colour of galaxy is more sensitive to the recent star formation history, thus being not capable of constraining its full star formation, especially that at early times. This is not in conflict with the conclusion that semi-analytical model over-predicts the mass of satellite galaxies.

 \item The galaxy bias, defined as the ratio of galaxy clustering to that of the dark matter distribution, has a general form in both models. That is, it is flat on large scales and rises at small scales without any strong transitional feature. The fact that this form appears independent of galaxy formation model supports the argument used by \cite{li2009}, who used this general shape to constrain cosmology parameter of $\sigma_{8}$. 

\end{itemize}

Overall, we find that the galaxy clustering is stronger affected by the model for galaxy formation than the adopted cosmology, thus it is currently impossible to accurately constrain the cosmological parameters. By comparing the predictions of the models in our paper, we find that the semi-analytical model  over-estimates the stellar mass of satellites. This over-prediction indicates that either the star formation efficiency in low-mass haloes at high redshifts is too high or that the tidal stripping of satellites is too inefficient. Furthermore, our results at $z=1$ shows that the stellar mass to halo mass relation is almost in place by $z=1$, suggesting that tidal stripping at $z<1$ should be weak. These results point to the fact that some other mechanisms, such as  QSO feedback, stronger tidal stripping at high redshifts, should be incorporated into the current galaxy formation models to remove the discrepancy between SAMs and observational data. These process are expected to be important at $z>1$ as QSO activities and galaxy interactions are much more common  in the past.

\section{Acknowledgements}
Xi Kang thanks Cheng Li for kindly providing the data and helpful discussions. XK is supported by the Bairen program of the Chinese Academy of Sciences, the foundation for the author of CAS excellent doctoral dissertation, and NSFC (No. 11073055). WPL acknowledges the support from Chinese National 973 project (No. 2007CB815401), NSFC project (No. 10873027, 10821302) and the Knowledge Innovation Program of the CAS (grant KJCX2-YW-T05). PJE acknowledges the support from CAS and SHAO. The simulation runs are supported by the Supercomputing center of CAS. 

\bibliographystyle{mn2e}
\bibliography{paper.bbl}


\label{lastpage}
\end{document}